\documentclass{PoS}

\title{Multi-wavelength X-ray/mid-infrared observations of GRS~1915+105}

\ShortTitle{X-ray/MIR observations of GRS 1915+105}

\author{\speaker{F. Rahoui}$^{ab}$, S. Chaty$^a$, J. Rodriguez$^a$, Y. Fuchs$^a$, F. Mirabel$^b$\\
\llap{$^a$}Laboratoire AIM, CEA/DSM - CNRS - Universit\'e Paris Diderot, Irfu/Service d'Astrophysique, B\^at. 709, CEA/Saclay, F-91191 Gif-sur-Yvette, France\\
\llap{$^b$}European Southern Observatory, Alonso de C\'ordova 3107, Vitacura, Santiago de Chile\\
        E-mail: \email{frahoui@cea.fr}, \email{chaty@cea.fr}, \email{jrodriguez@cea.fr}, \email{yfuchs@cea.fr}, \email{fmirabel@eso.org}
}

\abstract{We report preliminary results of mid-infrared (MIR) and X-ray observations of GRS~1915+105 that we carried out between 
2004 October 2 and 2006 June 5. Our main goals were to study its variability, to detect the presence of dust, and to 
investigate the possible links between MIR and X-ray emissions.

We performed photometric and spectroscopic observations of GRS~1915+105, using the IRAC photometer and the IRS spectrometer mounted on the \textit{Spitzer Space Telescope}. We completed our set of MIR data with quasi-simultaneous high-energy data obtained with \textit{RXTE} and \textit{INTEGRAL}. 

In the hard state, we detect PAH emission features in the MIR spectrum of GRS~1915+105, which prove the presence of dust in the system. The dust is confirmed by the detection in the hard state of a warm MIR excess in the broadband spectral energy distribution of GRS~1915+105. This excess cannot be explained by the MIR synchrotron emission from the compact jets as GRS~1915+105 was not detected at 15~GHz with the Ryle telescope. We also show that the MIR emission of GRS~1915+105 is strongly variable; it is likely correlated to the soft X-ray emission as it increases in the soft state. We suggest that, beside the dust emission,  part of the MIR excess in the soft state is non-thermal, and could be due either to free-free emission from an X-ray driven wind or X-ray reprocessing in the outer part of the accretion disc.}

\FullConference{VII Microquasar Workshop: Microquasars and Beyond\\
		 September 1-5 2008\\
		 Foca, Izmir, Turkey}

\begin{document}

\section{Introduction}

GRS~1915+105 was discovered in 1992 by the WATCH instrument on board the high-energy satellite \textit{GRANAT} (\cite{Castro1992}, \cite{Castro1994}). It became the first known Galactic source exhibiting superluminal radio jets when  \cite{Mirabel1994b} reported the VLA detection of 
radio outflows having an apparent superluminal velocity.  It is a Galactic low-mass X-ray binary (LMXB), composed of a 14$\pm$4~$M_{\odot}$ black hole (\cite{Greiner2001a}), accreting matter from a 0.81$\pm$0.53~$M_{\odot}$ K giant companion star (\cite{Greiner2001b}, \cite{Harlafatis2004}). The orbital period of the system is about 30.8$\pm$0.2~days (\cite{Neil2007}), and its distance is 9$\pm$3~kpc (\cite{Chapuis2004}).

GRS~1915+105 is strongly variable in the X-ray, near-infrared (NIR), and radio bands, on timescales from second to days.  Previous  multi-wavelength studies showed a strong connection between the accretion disc instabilities and plasma outflows (giant flares, compact jets, plasma bubbles) emitting at NIR and/or radio wavelengths through synchrotron processes (\cite{Fender1997}, \cite{Fender1998}, \cite{Eikenberry1998}, \cite{Mirabel1998}, \cite{Fender2000}, \cite{Fuchs2003a}). Moreover, \cite{Ueda2002} argue that about 20\%$-$30\% of the \textit{K} flux is due to reprocessing in the outer part of the accretion disc.

Few mid-infrared (MIR) observations of GRS~1915+105 were performed. Indeed, only \cite{Fuchs2003b} reported MIR photometry carried out with the ISOCAM camera on board the \textit{Infrared Space Observatory (ISO)} satellite. The authors observed GRS~1915+105 at two different epochs, corresponding to an intermediate soft state and a \textit{plateau} state of the source. They showed that, even if the flux uncertainties were high, the MIR flux of GRS~1915+105 was likely variable, and they reported a MIR excess, either due to the presence of dust, or to the synchrotron emission from compact jets, which is the explanation the authors favoured.

In this paper, we report the photometric and spectroscopic observations of GRS~1915+105 that we carried out with IRAC and IRS instruments on board the \textit{Spitzer Space Telescope}. We also used quasi-simultaneous \textit{RXTE} and \textit{INTEGRAL} data. Our goals were: 
\begin{itemize}
\item to confirm the variability of the MIR emission of GRS~1915+105.
\item to search for correlations between  its MIR and X-ray/radio emissions.
\item to investigate the presence of dust through MIR spectroscopy.
\item to study the evolution of the spectrum of GRS~1915+105 with respect to its high-energy states.
\end{itemize}
 
 We point out that in this work, for simplification, we will call hard state a phase during which the accretion disc is faint, and soft state a phase when it is brighter.
 \newline
 
 In Sect.~2, we give details on our observations and their reduction. In Sect.~3, we show the X-ray, MIR, and radio lightcurves of GRS~1915+105. 
 In Sect.~4, we explain how we assembled and fitted the broadband X-ray to MIR spectral energy distributions (SEDs) of GRS~1915+105, and we show 
 the MIR spectra in Sect.~5. We finally discuss the outcomes in Sect. 6.
 \clearpage
\section{Observations}

We performed, between 2004 October 2 and 2006 June 5, sixteen photometric 
and spectroscopic observations of GRS~1915+105 with the IRAC photometer 
and the IRS spectrometer (PI Fuchs), both mounted on the \textit{Spitzer Space Telescope}. 

We also carried out several observations of GRS~1915+105 with \textit{INTEGRAL} (PI Rodriguez), and 
several were quasi-simultaneous with our MIR observations.  Finally, the source is regularly 
monitored with \textit{RXTE} (PI Morgan, the data are immediately public). We browsed the archival data to search for pointings at  high energies quasi-simultaneous with our
MIR observations.  Several such pointings were found. \\

\subsection{IRAC photometry and IRS spectroscopy}

GRS~1915+105 was observed with IRAC at 3.6~$\mu$m, 4.5~$\mu$m, 5.8~$\mu$m, 
and 8~$\mu$m. We performed photometry on the post-Basic Calibration 
Data (post-BCD) using the software \textit{Starfinder}, part of the 
\textit{Scisoft} package from ESO, well-suited for point-source photometry 
in crowded fields. Post-BCD data are raw data on which the \textit{Spitzer} 
pipeline performs dark subtraction, multiplexer bleed correction, 
detector linearization, flatfielding, cosmic ray detection, flux calibration, 
pointing refinement, mosaicking, and coaddition. GRS~1915+105 was always 
detected, in all filters (see Table 1). 
\begin{table*}[h]
  $$ 
  \begin{array}{lccccc}
    \hline
    \textrm{Day}&3.6\,{\mu}\textrm{m}&4.5\,{\mu}\textrm{m}&5.8\,{\mu}\textrm{m}&8.0\,{\mu}\textrm{m}\\
    \hline
    \hline
    \textrm{Oct. 6 2004}&5.33 \pm 0.49&4.98\pm0.46&4.60\pm 0.49&3.11\pm0.44\\
    \hline
    \textrm{Oct. 31 2004}&4.89 \pm 0.47&4.49 \pm 0.45&4.09\pm 0.46&2.87\pm 0.43\\
    \hline
    \textrm{May 6 2005}&8.41 \pm 0.61&7.75 \pm 0.59&7.20\pm 0.60&5.10\pm 0.57\\
    \hline
    \textrm{May 10 2005}&9.95 \pm 0.66&9.40\pm 0.64&8.62 \pm 0.66&6.21\pm 0.63\\
    \hline
    \textrm{Sep. 23 2005}& 6.00  \pm 0.52&5.56\pm 0.49&5.19\pm 0.52&3.78\pm 0.50\\
    \hline
    \textrm{Nov. 2 2005}& 10.38 \pm 0.68&9.66\pm 0.65&8.96\pm 0.67&6.51\pm 0.64\\
    \hline
   \textrm{May 2 2006}& 4.70 \pm 0.46&4.54\pm 0.44&4.12\pm 0.46&2.94\pm 0.44\\
    \hline
    \textrm{Jun. 5 2006}& 4.96 \pm 0.47&4.69\pm 0.45&4.25 \pm 0.47&2.91\pm 0.44\\
    \hline
   \end{array}
  $$ 
  \caption{Summary of IRAC observations of GRS~1915+105. 
    We give the day of observation, and
    the MIR fluxes (in mJy). Uncertainties are given at 1$\sigma$.}
\end{table*}
\newline


We also performed spectroscopy of GRS~1915+105  with IRS using the SL2 (5.2~$\mu$m~$-$~7.7~$\mu$m) and 
SL1 (7.4~$\mu$m~$-$~14.5~$\mu$m) modules, with the IRS Peak-up option for a 
better pointing accuracy. BCD data were reduced following the standard 
procedure given in the IRS Data Handbook
\footnote{http://ssc.spitzer.caltech.edu/irs/dh/dh32.pdf}. The basics steps were 
sky subtraction, bad pixels correction, as well as 
extraction and calibration (wavelength and flux) of the spectra $-$ with the 
\textit{Spizer IRS Custom Extraction (Spice)} software $-$ for each nod. Spectra were 
then nod-averaged to improve the signal-to-noise ratio (SNR). 
GRS~1915+105 was always detected with SNRs good enough to allow the identification of spectroscopic features.

\subsection{\textit{RXTE} and \textit{INTEGRAL} observations}

The \textit{RXTE} data were reduced with the {\tt{HEASoft V6.5}} software
package, following the standard steps presented in the \textit{RXTE} cook 
book, and ABC guide. See e.g. \cite{Rodriguez2008a} and \cite{Rodriguez2008b}, for the details of 
good time filtering, background estimates, extraction of lightcurves and spectra, 
generation of response matrices .  Following these steps, we extracted lightcurves in the 
same three energy ranges as \cite{Belloni2000}, in order to classify a given observation in terms
of type of variability (\cite{Belloni2000}).  Spectra were then 
extracted from entire observations even from those showing clear spectral variations 
between the three spectral states, A, B, and C discussed in \cite{Belloni2000}.
\newline

The \textit{INTEGRAL} data from Rev~\#373 are presented in great 
details in \cite{Rodriguez2008a} and \cite{Rodriguez2008b} while Rev~\#431 is not discussed anywhere 
else. The data have been reduced with the {\tt{OSA v7.0}} software. Because 
of the wide field of view of \textit{INTEGRAL}, all bright sources have to be taken into 
account when extracting the data products. We then first ran the software 
until the production of images, that allowed us to identify those bright sources (defined 
as DETSIG$>$6) . All of them were considered during the extraction of spectra and 
light curves (see \cite{Rodriguez2008a} for the list of sources  in the case of Rev~\#373).

\section{Correlations}

\begin{figure}[h]
\begin{center}
\includegraphics[width=13cm]{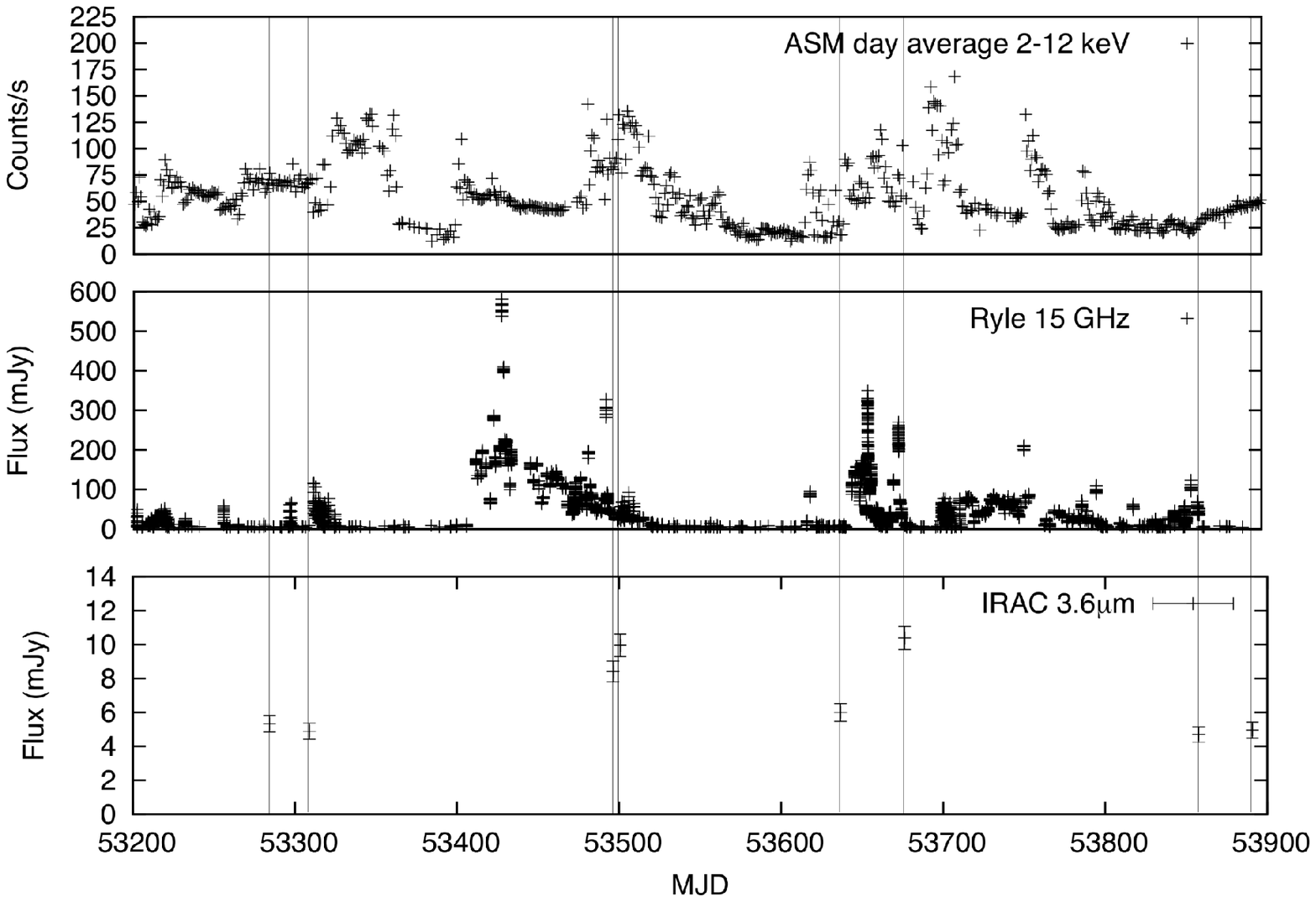}
\end{center}
\caption{Top: day average 2$-$12keV \textit{RXTE}/ASM lightcurve of GRS~1915+105.
\newline
Middle: Ryle Radio Telescope lightcurve of GRS~1915+105 at 15~GHz (see also \cite{Rodriguez2008a} and \cite{Rodriguez2008b}, courtesy G. G. Pooley).
\newline
Bottom: IRAC lightcurve of GRS~1915+105 at 3.6~$\mu$m.
}
\end{figure}

We first searched for X-ray/MIR/radio correlations by comparing the GRS~1915+105 data in these domains. 
Figure~1 displays the day average 2$-$12~keV \textit{RXTE}/ASM lightcurve, as well as the 15~GHz Ryle Radio Telescope (PI Pooley) and the 3.6~$\mu$m IRAC lightcurves between MJD~53200 and MJD~53900. 

We do not have enough MIR data to  search for accurate correlations, and they, moreover, are integrated over a long exposure time considering the very rapid X-ray to radio flux variations GRS~1915+105 can exhibit. Nevertheless, the comparison between IRAC and ASM lightcurves shows that the strongly variable MIR flux of GRS~1915+105 increases when the accretion disc enters into a very unstable state, while it is low and steady when GRS~1915+105 is in a stable state. This behaviour seems to indicate an X-ray/MIR correlation. On the contrary, the comparison between radio and MIR lightcurves does not allow any conclusion as the MIR flux of GRS~1915+105 does not seem to follow any parallel evolution to its radio emission.

\section{SED}

For each quasi-simultaneous IRAC and {\textit{RXTE}/\textit{INTEGRAL} data, 
we built the X-ray to MIR broadband SEDs of GRS~1915+105, for different states of the source. 
We then fitted each of them, respecting the following procedure: 
\begin{itemize}
\item the high-energy fluxes were fitted with a model consisting of a multicolor blackbody (accretion disc), a gaussian (iron feature at 6.5~keV), and a powerlaw component, all modified by absorption. The absorbing column density was frozen at 5.3~$\times 10^{22}$~atoms~cm$^{-{2}}$, following \cite{Rodriguez2008b}. 
\item IR fluxes were dereddened, from a 19.6$\pm$1.7 visual absorption (\cite{Chapuis2004}), 
using the extinction law given in \cite{Indebetouw2005}.
\item the dereddened X-ray to MIR broadband SED was finally fitted using a model 
including the previously found accretion disc, gaussian, and powerlaw, as well as a stellar 
blackbody for the companion star. 
\end{itemize}
Each time, we tried to reproduce the infrared emission of GRS~1915+105 adding the stellar blackbody only, but we were never able to correctly 
fit the SED, neither in the hard nor in the soft state, since a MIR excess was always present. 
The companion star of GRS~1915+105 being a KM giant\cite{Greiner2001b}, we then froze the stellar blackbody temperature 
and radius to 4000~K and 25R$_{\odot}$ respectively, and we added an extra blackbody component accounting for this MIR 
excess. Figure~2 displays two fitted broadband SEDs corresponding to the hard state (faint disc) and the soft state (bright disc) of GRS~1915+105, and Table~2 lists the best-fit parameters for the disc and the MIR excess, for each SED.

Figure~2 and the parameters given in Table~2 confirm the X-ray/MIR correlation suggested in Figure~1. Indeed, for all the SEDs we fitted 
using the same model with quasi-simultaneous X-ray to MIR data, the MIR flux increases with the accretion disc flux. 
Moreover, whether in the hard or soft state, the MIR excess best-fit temperature is constant, while its luminosity increases with the brightness of the accretion disc. 

To summarize, the conclusions of the SEDs fitting are: 
\begin{itemize}
\item GRS~1915+105 exhibits a MIR excess,  in the hard and soft states, which suggests the existence of a warm  component in the system.
\item the MIR emission is correlated to the soft X-ray emission from the disc.
\item the MIR excess temperature is constant but its luminosity increases with the brightness of the disc, which suggests either a heating of another component by the X-ray emission, or a non-thermal origin of this increase.
\end{itemize}

\begin{figure}[h]
\includegraphics[width=8cm]{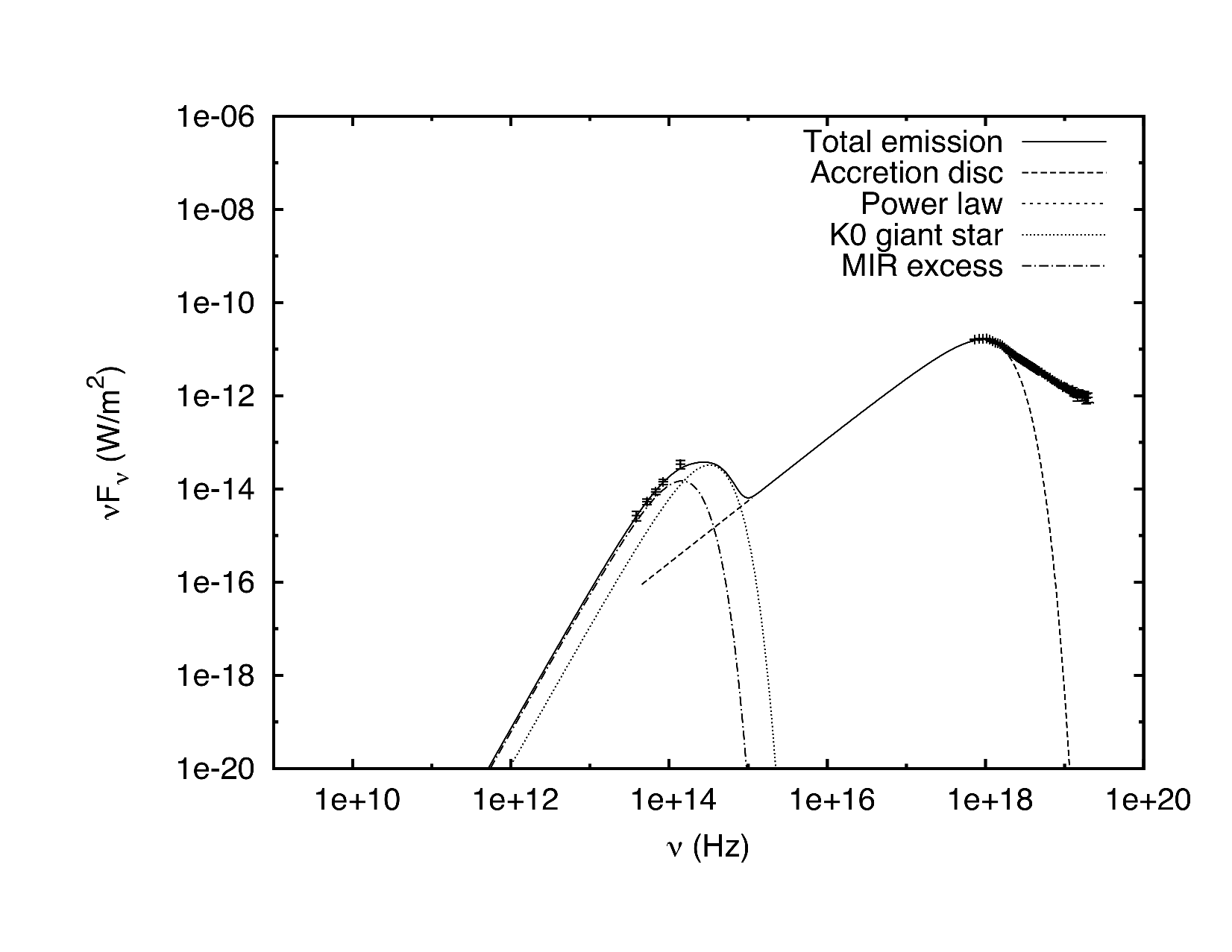}
\includegraphics[width=8cm]{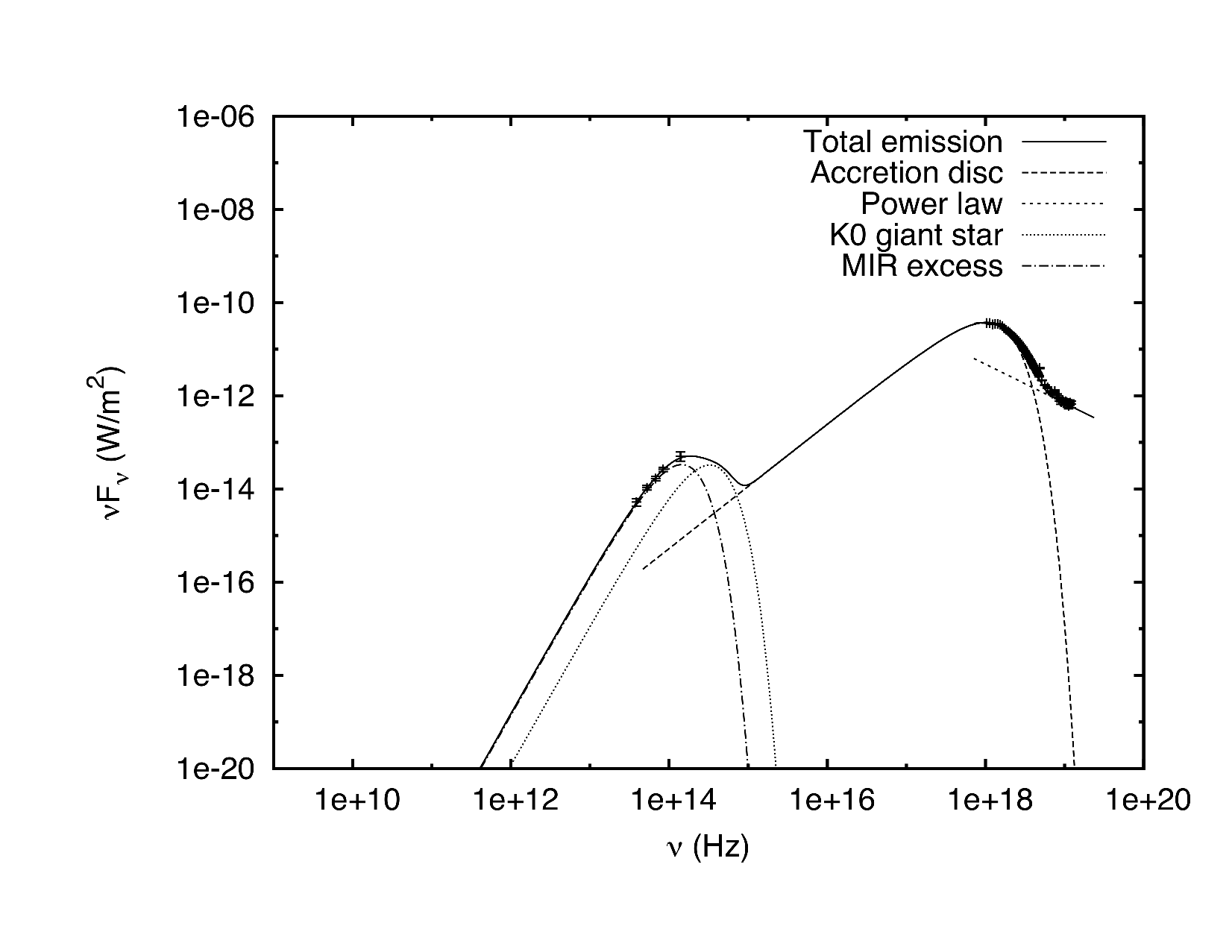}
\caption{Left: X-ray to MIR broadband SED of GRS~1915+105 for the hard state of the source (MJD~53284.176).
\newline
Right: X-ray to MIR broadband SED of GRS~1915+105 for the soft state of the source (MJD53500.571).}
\end{figure}

\begin{table*}[h]
  $$ 
  \begin{array}{ccc}
    \hline
    \textrm{MJD}&53284.176&53500.571\\
    \hline
    \hline
    F_{\rm 3.6\mu m}/\textrm{mJy}&5.33\pm0.59&9.95\pm0.66\\
    \hline
    F_{\rm disc}/\textrm{erg}~\textrm{cm}^{-{2}}~\textrm{s}^{-{1}}&1.7\times 10^{-{8}}&4.4\times 10^{-{8}}\\
    \hline
    T_{\rm exc}/\textrm{K}&1188\pm142&1168\pm123\\
    \hline
    L_{\rm exc }/\textrm{erg}~\textrm{s}^{-{1}}&2.8\pm1.0&5.3\pm1.0\\
    \hline
 \end{array}
  $$ 
  \caption{Best-fit parameters for MJD~53284.176 and 53500.571. We give the accretion disc flux, and the MIR excess 
  temperature and luminosity. We also give the corresponding MIR flux at 3.6 $\mu$m.}
\end{table*} 

~
\section{MIR spectra}

We showed in the previous section that GRS~1915+105 exhibited a MIR excess when the source was 
in the hard or soft state.  To understand the origin of this excess, we then compared several MIR spectra that we obtained for different high-energy states. 
Figure~3 displays two of them, obtained when GRS~1915+105 was in the hard state (faint disc) and in the soft state (bright disc). In the hard state, we clearly detect three PAH emission features at 6.2~$\mu$m, 7.7~$\mu$m, and 11.3~$\mu$m. PAHs are strong tracers of dust,  
and their presence in emission in the MIR spectrum of GRS~1915+105 strongly suggests the existence of a heated dust component in the system. 

Moreover, in the soft state, the MIR continuum below 8~$\mu$m increases, which confirms the X-ray/MIR correlation. Note that  the PAH features disappeared, which could prove that the dust was destroyed or expelled from the system. We point out that the PAH feature at 11.3~$\mu$m also disappeared although there is no increase of the MIR continuum beyond 10~$\mu$m, which excludes that the PAH features originate from the photoionised ISM, their disappearance being due to the pollution by the continuum.

\begin{figure}[h]
\includegraphics[width=8cm]{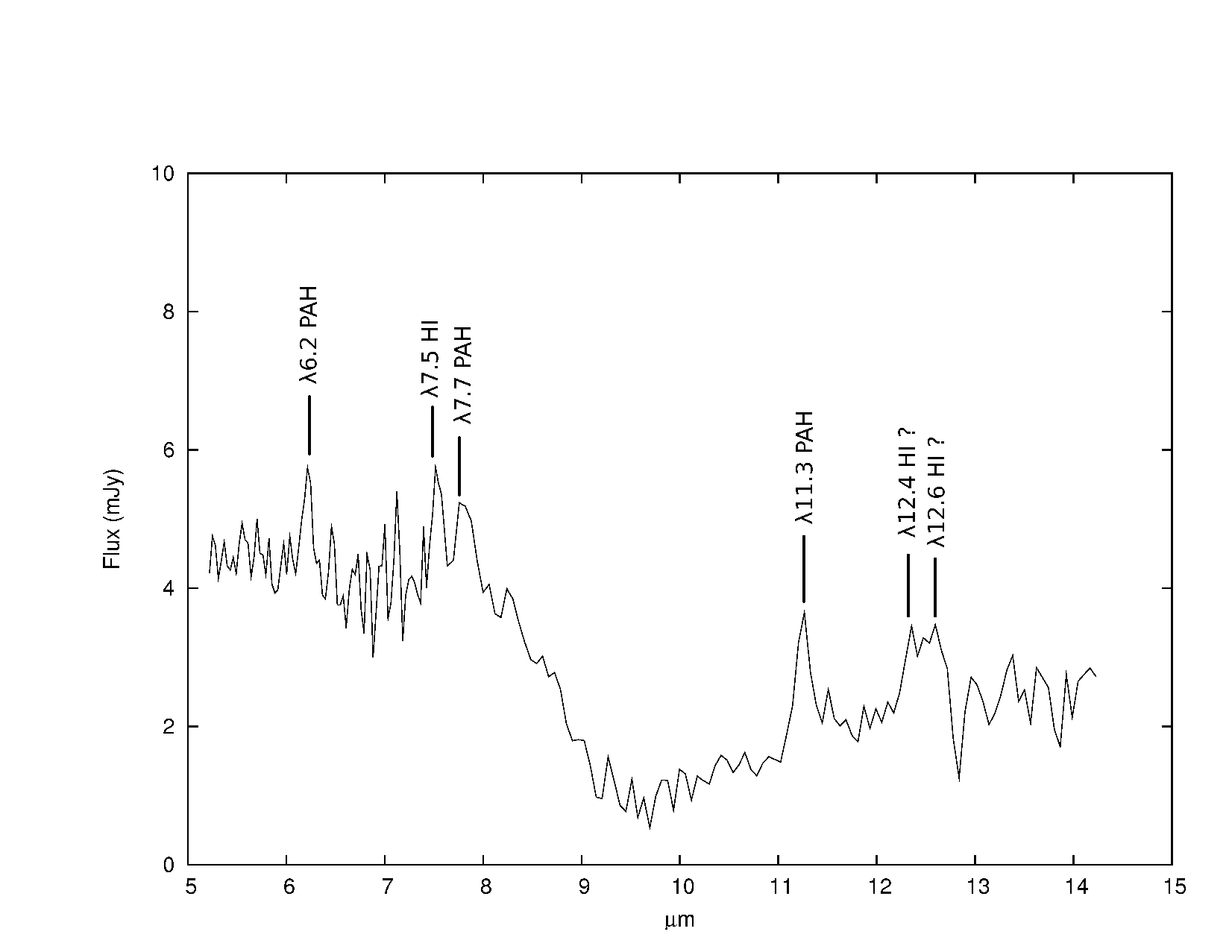}
\includegraphics[width=8cm]{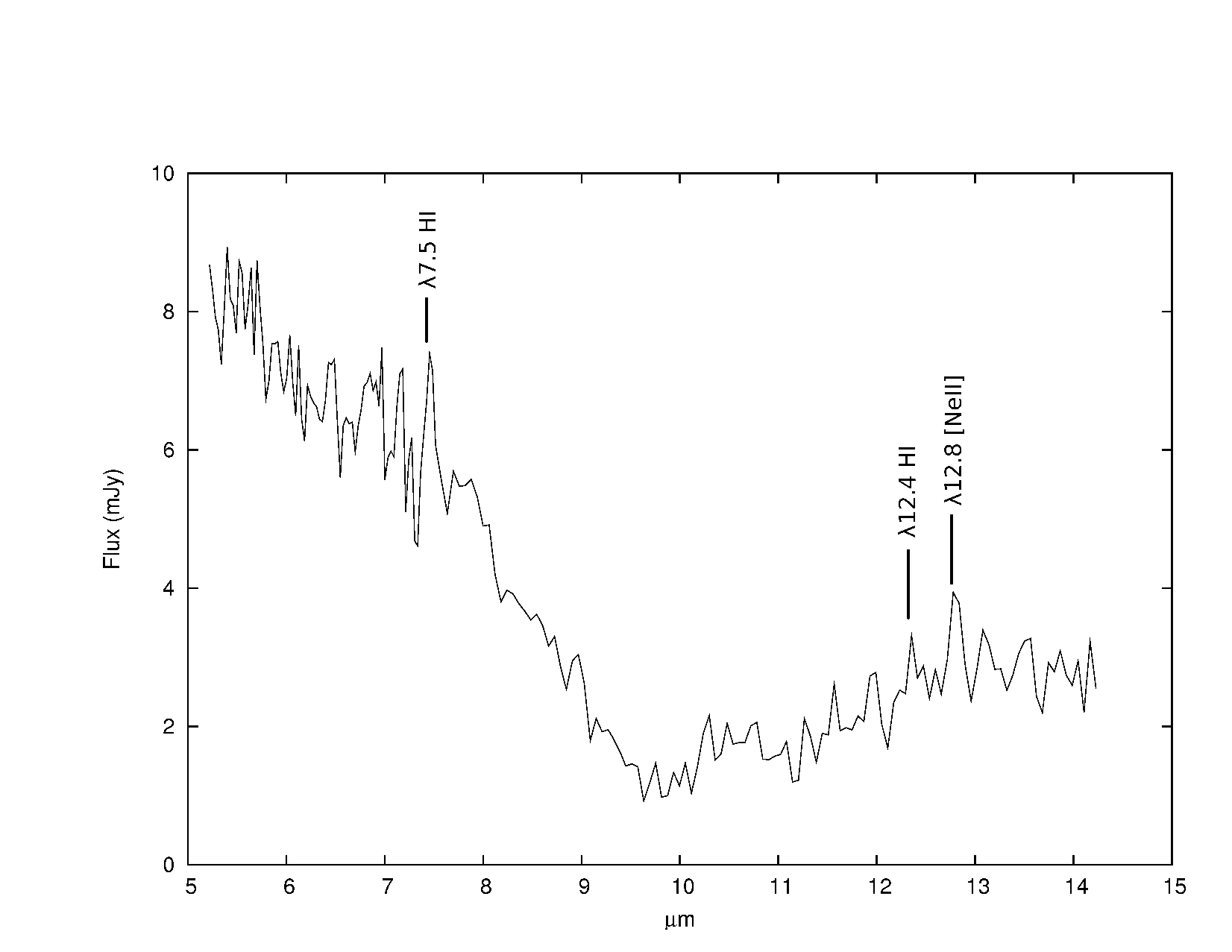}
\caption{Left: MIR spectrum of GRS~1915+105 for the hard state of the source (MJD~53280.243).
\newline
Right:MIR spectrum of GRS~1915+105 for the soft state of the source (MJD53511.697).
}
\end{figure}

\section{Conclusion}

The MIR excess and the detection of PAH features in the hard state of GRS~1915+105 is typical of a warm circumbinary material. 
The presence of dust around GRS~1915+105 was already suggested by \cite{Mirabel1996} and \cite{Lee2002}, and \cite{Muno2006} showed that such 
component was probably responsible for the MIR excess detected in the emission of A0620$-$00 and XTE~J1118$+$338. 
The origin of the dust around LMXBs is still question of debate, but it could originate from the remains of a supernova fallback disc, or from the dusty giant companion star. 
Moreover, we show that the MIR emission of GRS~1915+105 was strongly variable and correlated to the soft X-ray emission from the accretion disc, as the MIR flux and continuum below 8~$\mu$m increase in the soft state. The reasons of such correlations could be either heating of the dust by the X-ray emission, X-ray reprocessing of hard X-ray in the outer part of the disc, free-free emission of a X-ray driven wind, or synchrotron emission from a compact jet. If \cite{Migliari2006} and \cite{Migliari2007} favour the latter explanation for 4U~0614$+$091 and GRO~J1655$-$40, we exclude this possibility concerning GRS~1915+105 as synchrotron emission from a compact jet would lead to an increase of all the MIR continuum, which we do not detect beyond 10~$\mu$m. Heating of the dust is also unlikely, as the PAH features disappear in the soft state. The MIR variability of GRS~1915+105 then likely originates from non-thermal processes. Whether it is due to the X-ray reprocessing or free-free emission is still a work in progress, and we need further observations and modelling to conclude.

\end{document}